\documentclass{ieeeaccess}
\usepackage{amsmath,amssymb,amsfonts}
\usepackage{graphicx}
\usepackage{textcomp}
\usepackage{comment}

\usepackage{etoolbox}

\def\thevol{XXXX}
\def\myyear{XXXX}
\makeatletter
\patchcmd{\@evenfoot}{2016}{\myyear}{}{}
\patchcmd{\@oddfoot}{2016}{\myyear}{}{}
\makeatother

\def\BibTeX{{\rm B\kern-.05em{\sc i\kern-.025em b}\kern-.08em
    T\kern-.1667em\lower.7ex\hbox{E}\kern-.125emX}}

\usepackage{cite}
\usepackage{psfrag}
\usepackage{url}
\usepackage{stfloats}
\usepackage{amsmath}
\usepackage{array}
\usepackage{epsfig}
\usepackage{amssymb}
\usepackage{color}
\usepackage{subfig}
\usepackage{setspace}
\usepackage{caption}
\usepackage{booktabs}
\usepackage{ dsfont }
\usepackage{algpseudocode}

\usepackage[square,numbers,sort&compress]{natbib}
\usepackage{algorithm}  
\usepackage{amsmath,epsfig,amssymb,algpseudocode,amsthm,cite,url}
\usepackage{comment}

\usepackage{multirow}
\DeclareMathOperator*{\argmax}{argmax}

\graphicspath{{"./pictures/"}}

\newcommand\blfootnote[1]{%
	\begingroup
	\renewcommand\thefootnote{}\footnote{#1}%
	\addtocounter{footnote}{-1}%
	\endgroup
}

\ifCLASSINFOpdf
\graphicspath{{./pictures/}}
\else
\usepackage{enumerate}
\usepackage{amsfonts,amssymb}
\fi

\begin{document}
\history{Date of publication xxxx 00, 0000, date of current version xxxx 00, 0000.}
\doi{xx.xxxx/ACCESS.2020.DOI}

\title{A Deep Reinforcement Learning Approach to Efficient Drone Mobility Support}
\author{\uppercase{Yun Chen}\authorrefmark{1}, \IEEEmembership{Student Member, IEEE},
\uppercase{Xingqin Lin}\authorrefmark{2}, \uppercase{Talha A. Khan}\authorrefmark{2}, and \uppercase{Mohammad Mozaffari}\authorrefmark{2}\vspace{0.2cm}}
\address[1]{Department of Electrical and Computer Engineering, The University of Texas at Austin, Austin, TX 78712 USA (e-mail: yunchen@utexas.edu)}
\address[2]{Ericsson Research Silicon Valley, Santa Clara, 
CA 95054 USA (e-mail: \{xingqin.lin, talha.khan, mohammad.mozaffari\}@ericsson.com)}

\markboth
{Y. Chen \headeretal: A Deep Reinforcement Learning Approach to Efficient Drone Mobility Support}
{Y. Chen \headeretal: A Deep Reinforcement Learning Approach to Efficient Drone Mobility Support}

\corresp{Corresponding author: Talha A. Khan (e-mail: talha.khan@ericsson.com). 
}


\begin{abstract}	
	The growing deployment of drones in a myriad of applications relies on seamless and reliable wireless connectivity for safe control and operation of drones. 
	Cellular technology is a key enabler for providing essential wireless services to flying drones in the sky. 
	Existing cellular networks targeting terrestrial usage can support the initial deployment of low-altitude drone users, but there are challenges such as mobility support.
	In this paper, we propose a novel handover framework for providing efficient mobility support and reliable wireless connectivity to drones served by a terrestrial cellular network. 
	Using tools from deep reinforcement learning,  
	we develop a deep Q-learning algorithm to dynamically optimize handover decisions to ensure robust connectivity for drone users. 
	Simulation results show that the proposed framework significantly reduces the number of handovers at the expense of a small loss in signal strength relative to the baseline case where a drone always connect to a base station that provides the strongest received signal strength.  
\end{abstract}


\titlepgskip=-1pt

\maketitle

\section{Introduction}
\label{sec:intro}
\blfootnote {This work was initiated when Yun Chen was with Ericsson Research.}
Due to the unique advantages of drones such as swift mobility and low-cost operation, their applications are rapidly growing from item delivery and traffic management to asset inspection and aerial imaging \cite{saad2020wireless, TutorialMO, fotouhi2019survey, B5GUAV}. 
Realizing the true potential of drone technology hinges on ensuring seamless wireless connectivity to drones.
Cellular technology is well-suited for providing connectivity services to drones thanks to its reliability, flexibility and ubiquity.  
Several efforts are underway to develop cellular-assisted solutions, leveraging Long-Term Evolution (LTE) and the fifth-generation (5G) New Radio (NR), for supporting efficient drone operations in the sky \cite{yang2018telecom, Sky}. 
To better understand the potential of cellular networks for low-altitude drones, the third-generation partnership project (3GPP) has been studying and developing new features for enhanced mobile services for drones acting as user equipments (UEs) \cite{3GPPTR, 3GPP22825, muruganathan2018overview}. To further meet the needs of 5G connectivity of drones, a new 3GPP activity is planned to devise new key performance indicators and identify communication needs of a drone with a 3GPP subscription. In addition, 3GPP is evolving 5G NR to support non-terrestrial networks \cite{3GPP38811, 3GPP38821, lin20195g}. It is expected that the more flexible and powerful NR air interface will deliver more efficient and effective connectivity solutions for wide-scale drone deployments \cite{lin20195gnr}.

While the low-altitude sky is within reach of existing cellular networks, 
enabling robust and uninterrupted services to aerial vehicles such as drones poses several challenges. 
We next review some of the key technical challenges in serving drone UEs using existing cellular networks. 
First, terrestrial cellular networks are primarily designed for serving ground UEs and usually use down-tilted base station (BS) antennas. This means that drone UEs are mainly served by the side lobes of the BS antennas and may face coverage holes in the sky due to nulls in the antenna pattern \cite{MobilieDrones}. 
Second, the drone-BS communication channels have high line-of-sight probabilities. As a result, a drone UE may generate more uplink interference to the neighbouring cells and experience more interference in the downlink as signals from many neighboring cells may reach the drone with strong power levels. The strong interference, if not properly managed, may degrade link quality of both ground UEs and drone UEs. 
Third, the high speed and three-dimensional motion of drones make handover (HO) management more cumbersome compared to ground UEs. 
In a network with multiple BSs (serving multiple cells), these challenges further compound the drone-BS association rules. This is because the coverage space formed by the strongest BSs is no longer contiguous but rather fragmented \cite{MobilieDrones}. This, in turn, can trigger frequent HOs leading to undesirable outcomes such as radio link failures, ping-pong HOs, and large signaling overheads. 
This motivates the need for an efficient HO mechanism that can provide a robust drone mobility support in the sky \cite{chen2019efficient}.

\subsection {Related Work}

The support of mobility is a fundamental aspect of wireless networks \cite{camp2002survey, akyildiz1999mobility}. Mobility management is particularly an essential and complex task in emerging cellular networks with small and irregular cells \cite{lin2013towards, andrews2013seven}. There has been a recent surge of interest in applying machine learning techniques to mobility management in cellular networks. In \cite{wickramasuriya2017base}, a recurrent neural network (RNN) was trained using sequences of received signal strength values to perform BS association. In \cite{mismar2018partially}, a supervised machine learning algorithm was proposed to improve the success rate in the handover between sub-6 GHz LTE and millimeter-wave bands. In \cite{yajnanarayana20195g}, a HO optimization scheme based on reinforcement learning (RL) was proposed for terrestrial UEs in a 5G cellular network. In \cite{alkhateeb2018machine}, a HO scheme based on deep learning was proposed to improve the reliability and latency in terrestrial  millimeter-wave mobile networks.

In 3GPP Release 15, a study was conducted to analyze the potential of LTE for providing connectivity to drone UEs \cite{3GPPTR}. This study identified mobility support for drones as one of the key areas that can be improved to enhance the capability of LTE networks for serving drone UEs. In \cite{stanczak2018mobility}, an overview of the major mobility challenges associated with supporting drone connectivity in LTE networks was presented. In \cite{MobilityS}, the performance of a cellular-connected drone network was analyzed in terms of radio link failures and number of HOs. In \cite{challita2019interference}, an interference-aware drone path planning scheme was proposed and the formulated problem was solved using a deep RL algorithm based on echo state network. In \cite{fakhreddine2019handover}, HO measurements were reported for an aerial drone connected to an LTE network in a suburban environment.
The results showed how HO frequency increases with increasing flight altitude, based on which the authors suggested that enhanced HO techniques would be required for a better support of drone connectivity.

While prior work has studied various mobility challenges pertaining to drone communications, efficient HO optimization for drone UEs (as motivated in Section \ref{sec:intro})
has received little attention. To this end, in our recent work \cite{chen2019efficient}, a HO mechanism based on Q-learning was proposed for a cellular-connected drone network. It was shown that a significant reduction in the number of HOs is attained while maintaining reliable connectivity. The promising results have inspired further work such as \cite{chowdhury2020mobility} that adopted a similar approach for drone mobility management by tuning the down-tilt angles of BSs.

\subsection{Contributions}

The aim of our work is to find an efficient HO mechanism which accounts for the mobility challenges faced by drone UEs in a terrestrial cellular network optimized for serving devices on the ground. In this paper, we present the second part of our work on using RL to improve drone mobility support, completing the first part of our work presented in our recent paper \cite{chen2019efficient}. Despite the encouraging results in \cite{chen2019efficient}, the tabular Q-learning framework adopted in \cite{chen2019efficient} may have some disadvantages. 
First, the algorithm may entail substantial storage requirements when the state space is large. For example, this is the case with long flying routes having numerous waypoints where the drone needs to make HO decisions. This problem will be further exacerbated when there is a large pool of candidate cells to choose from.
Second, the Q-learning approach adopted in \cite{chen2019efficient} can only be used for discrete states, which implies that the proposed scheme therein can help make HO decisions only at predefined waypoints rather than at arbitrary points along the route. 
These disadvantages are addressed in this paper by using tools from deep RL \cite{sutton1998introduction, li2018deep}. 

In this paper, we propose a deep Q-network (DQN) based optimization mechanism for a cellular-connected drone system to ensure robust connectivity for drone UEs. With the use of deep RL tools, HO decisions are dynamically optimized using a deep neural network to provide an efficient mobility support for drones. In the proposed framework, we 
leverage reference signal received power (RSRP) data and a drone's flight information 
to learn effective HO rules for seamless drone connectivity while accounting for HO signaling overhead. Furthermore, our results showcase the inherent
interplay between the number of HOs and the serving cell RSRP in the considered cellular system. We also compare our results to those reported in \cite{chen2019efficient} that adopted an approach based on Q-learning.

The rest of this paper is organized as follows. The system model is described in Section II. A brief background of deep RL in the context of our work is introduced in Section \ref{Background of RL}. A DQN-based HO scheme is presented in Section \ref{Model and Algorithms}. The simulation results are provided in Section \ref{Experiments and Results}. Section \ref{Conclusion} concludes the paper.

\section{System Model}
We consider a terrestrial cellular network with down-tilted BS antennas. 
Traditionally, such a network mainly targets serving users on the ground. In this work, we assume that it also serves drone UEs flying in the sky. Each drone UE moves along a two-dimensional (2D) trajectory at a fixed altitude. One of the main goals of the cellular network is to provide fast and seamless HO from one cell (a source cell) to another (a target cell). Due to 
its
high mobility nature, 
a drone UE may experience multiple HO events, resulting in 
frequent switching of 
serving cells along its trajectory.

\begin{figure}
	\centering
	\includegraphics[width=8.2cm]{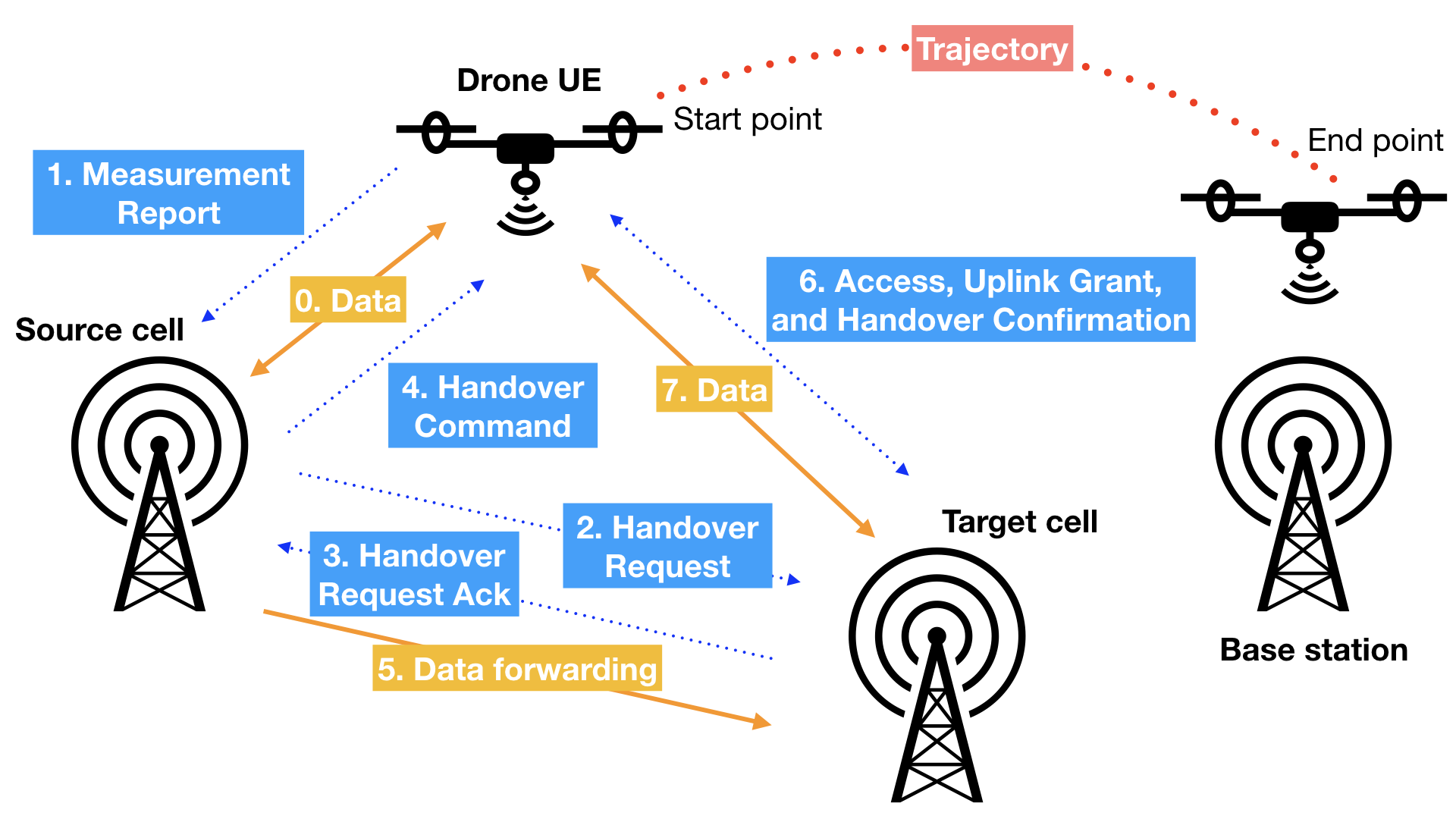}
	\caption{Illustration of network model and handover procedure.}
	\label{fig:1}
\end{figure}

Fig. \ref{fig:1} illustrates a typical network-controlled HO procedure with drone UE assistance. In the source cell, the drone UE is configured with measurement reporting. That is, it performs measurements such as RSRP to assess its radio link quality and reports the results to the network. For mobility management, the drone UE measures the signal quality of neighbor cells in addition to its serving cell. The network may use the reported measurement results for making a HO decision. If a HO is deemed necessary, the source cell sends a HO request to the target cell. After receiving an acknowledgment of the HO request from the target cell, the source cell can send a HO command to the drone UE. Meanwhile, the source cell may carry out data forwarding to the target cell. Upon receiving the HO command, the drone UE can initiate the random access procedure towards the target cell, receive uplink grant, and send the HO confirmation message. Once the HO procedure is completed, the drone UE can resume data communication with the network through the target cell (which becomes the serving cell upon HO completion).

We assume that the drone trajectory is fixed and known to the network. We consider predefined locations (or waypoints) along the trajectory for making HO decisions. For each such location, it is first decided whether a HO is needed or not. In case a HO is needed, a target cell is further decided. The HO decisions may depend on various factors such as BS distribution, drone mobility profile including speed and flight trajectory, and propagation environment, among others. 

We consider a baseline HO strategy purely based on RSRP measurements where  the drone UE is assumed to always connect to the cell which provides the largest RSRP. While selecting the strongest cell is indeed appealing from a radio signal strength perspective, a HO decision solely based on the largest RSRP at the waypoint is often short-sighted as it may trigger many subsequent HO events during the flight. Further, the considered baseline HO strategy may cause frequent ping-pong HO events and radio link failures. This is because the signal strength can fluctuate rapidly along the drone trajectory in a cellular network with down-tilted BS antennas. Also, there can be a service interruption 
during the time interval
when the drone UE receives a HO command from the source cell until the target cell receives the HO confirmation from the drone UE.
In short, HO is a costly procedure, hence the number of HO events along the flight trajectory needs to be minimized while maintaining the desired radio link quality.

In this work, we use RSRP as a proxy for radio link reliability and the number of HO events as a measure of HO cost which may include signaling overhead, potential radio link failure, and service interruption time associated with the HO procedure. Intuitively, a desirable HO mechanism will lead to sufficiently large RSRP values while incurring only a modest number of HO events along a flight trajectory. To this end, we propose a deep RL-based framework to determine the optimal sequential HO decisions 
to achieve reliable connectivity while accounting for the HO costs. In this framework, we consider two key factors in the objective function: 1) the serving cell RSRP values, and 2) the cost (or penalty) for performing a HO. To reflect the impact of these factors in the HO decisions, we define $w_{RSRP}$ and $w_{HO}$ as the weights of the serving cell RSRP and the HO cost, respectively. From a design perspective, 
adjusting the weights $w_{RSRP}$ and $w_{HO}$ can help strike a balance between maximizing the serving cell RSRP values and minimizing the number of HO events.

\begin{algorithm}[!t]  
\caption{Training process for HO scheme using DQN.}  
\label{algorithm_Q}  
{\small
	\begin{algorithmic}[1]   
		\State Initialize input parameters: 
		\Statex Drone trajectory $\mathcal{R}=\{\mathcal{P}_i|i=0,1,...,l-1\}$;
		\Statex Firstly connected cell $c_0$ at  $\mathcal{P}_0$ (usually the strongest cell);
		\Statex$C_{k_{s}}\leftarrow k$ strongest cells at waypoint $\mathcal{P}_1$;
		\Statex Action-value function $Q(s,a;w)$ with random weights $w$;
		\Statex Target action-value $\hat{Q}(s,a;w')$ with weights $w' = w$;
		\Statex Number of episodes $n$; Number of training steps $\tau$;
		\Statex Replay memory $\boldsymbol{\beta}$; Minibatch size $m$;
		\Statex Threshold $\mathcal{T}_h$ for initiating Q-value iteration;
		\Statex DQN update cycle $\mathcal{T}_c$; 
		\Statex Discount factor $\lambda$, Exploration coefficient $\epsilon$;
\item[]
		\While{episode $< n$}
		\State Initialize state $s:[x_0, y_0, \theta_0, c_0]$;
		\State step $\leftarrow 0$;
		\While{step $<\mathcal{T}$}	
		\If{Current position is the termination $(x_{l-1},y_{l-1})$}
		\State Randomly pick a waypoint from $\mathcal{R}$ and update $s$;
		\EndIf
		\item[]
		\If{$\epsilon<=$ UniformRandom[0,1] }
		\State $a_s$ $\gets$ Randomly pick an index from $C_{k_s}$;
		\Else
		\State $a_s$ $\gets$ $\argmax\limits_{a} Q(s,a;w)$;
		\EndIf
		\State Execute $a_s$, transit to $s'$ and get reward $R_{s,a_s}$;
		\State Store transition $(s, a_s, R_{s,a_s}, s')$ into batch $\boldsymbol{\beta}$;
		\State Update current state: $s\leftarrow s'$;
		\State Update $C_{k_{s}}$;
		\item[]
				\If{size($\boldsymbol{\beta}$) $>= m$}
		\State Sample a random minibatch $\boldsymbol{\beta}_m$ from $\boldsymbol{\beta}$;
		\State Calculate $\mathbb{Q}(\boldsymbol{\beta}_{m,S};w) \in \mathds{R}^{m\times k}$ for states $\boldsymbol{\beta}_{m,S}$;
	
		\If{step / $\mathcal{T} < \mathcal{T}_h$}
\State	Update $\mathbb{Q}(\boldsymbol{\beta}_{m,S};w)$: 
	$\forall\, i\in\mathcal{I}_m, j\in\mathcal{I}_k,$ \Statex \hskip7em $Q(\boldsymbol{\beta}_{m,S}(i),j;w) \leftarrow R_{\beta_{m,S}(i),j} $
		\Else
\State	Update $\mathbb{Q}(\boldsymbol{\beta}_{m,S};w)$: 
	$\forall\, i\in\mathcal{I}_m, j\in\mathcal{I}_k,$ \Statex $Q(\boldsymbol{\beta}_{m,S}(i),j;w) \leftarrow R_{\beta_{m,S}(i),j}+\lambda \max\limits_{a'}\hat{Q}(\boldsymbol{\beta}_{m,S}(i)',a';w') $

		\EndIf
		\If{$\text{mod}\left(\text{step}, \mathcal{T}_c\right) =0$}
		\State $w' \leftarrow w$;
		\EndIf  
		
		\Else
		\State Continue;
		\EndIf
		\State Perform a gradient descent to minimize the loss function (Eq.~\ref{loss_func}) with respect to parameter $w$;
		\State step $\leftarrow$ step$+1$;
		\EndWhile
		\State episode $\leftarrow$ episode$+1$;
		\EndWhile
	\end{algorithmic}
}
\end{algorithm}

\section{Background of Deep Reinforcement Learning}\label{Background of RL}

As a subfield of machine learning, RL addresses the problem of automatic learning of optimal decisions over time \cite{sutton1998introduction}. A RL problem is often described by using a Markov decision process characterized by a tuple $(\mathcal{S}, \mathcal{A}, P, \lambda,{R})$, where $\mathcal{S}$ denotes the set of states, $\mathcal{A}$ denotes the set of actions, $P$ denotes the state transition probabilities, $\lambda\in[0,1)$ is a discounting factor that penalizes the future rewards, and $R$ denotes the reward function. In RL, an agent interacts with an environment by taking actions based on observations and the anticipated future rewards. Specifically, the agent can stay in a state $s\in\mathcal{S}$ of an environment, take an action $a\in\mathcal{A}$ in the environment to switch from one state to another governed by the state transition probabilities, and in turn it receives a reward $R$ as feedback. The RL problem is solved by obtaining an optimal policy $\pi^*$ which provides the guideline on the optimal action to take in each state such that the expected sum of discounted rewards is maximized.

Q-learning, as adopted in our previous work \cite{chen2019efficient}, is one of the most promising algorithms for solving RL problems \cite{Q-learning}. Let us denote by $Q^{\pi}(s,a)$ the Q-value (or action-value) of a state-action pair $(s, a)$ under a policy $\pi$. Formally, $Q^{\pi}(s,a) = \mathbb{E}[G_t|S_t=s, A_t=a]$, where $G_t = \sum_{k=0}^\infty \lambda^k R_{t+k+1}$ is the return at time $t$. So, $Q^{\pi}(s,a)$ is the expected sum of discounted rewards when the agent takes an action $a$ in state $s$ and chooses actions according to the policy $\pi$ thereafter. The optimal policy $\pi^*$ achieves optimal value function: $\pi^* = \arg \max_{\pi} Q^{\pi}(s,a)$. Thus, by computing the optimal Q-values $Q^{*}(s,a)$, one can derive the optimal policy that chooses the action with the highest Q-value at each state. The optimal Q-values can be computed by iterative algorithms. With a slight abuse of notation, we use $Q_{t}(s,a)$ to denote the Q-value at time $t$ during the iterative process. When the agent performs an action $a$ in a state $s$ at time $t$, it receives a reward $R_{t+1}$ and switches to state $s'$. The Q-value iteration process is given by
\begin{align}
\label{Q_iteration}
Q_{t+1}(s,a)&\leftarrow (1-\alpha)~ Q_t(s,a)\nonumber\\
&\qquad+
\alpha\left[R_{t+1}+\lambda \max \limits_{a'\in\mathcal{A}}Q_t(s',a')\right],
\end{align}
where $\alpha$ is the learning rate. It can be shown that $Q_t$ approaches $Q^{*}$ when $t\rightarrow \infty$. This method is known as tabular Q-learning.

The aforementioned Q-learning method may be difficult to use in problems with large state space, as the number of Q-values grows exponentially with state space variables. A nonlinear representation that maps both state and action onto a value can be used to address this issue. Neural networks are universal approximators and have drawn significant interest from the machine learning community. It has been shown that the depth of a deep neural network can lead to an exponential reduction in the number of neurons required \cite{delalleau2011shallow}. Thus, using a deep neural network for value function approximation is a promising option. Let us denote by $Q(s, a; w)$ the approximated Q-value function with parameters $w$. A DQN \cite{mnih2013playing} aims to train a neural network with parameters $w$ to minimize the loss function $L (w)$ where
\begin{align}
\label{loss_func}
L (w) = \mathbb{E} \left[ \left( R + \lambda \max \limits_{a'\in\mathcal{A}}Q_t(s',a'; w) -  Q(s, a; w) \right)^2 \right].
\end{align}

\section{Deep RL-Based HO Optimization Framework}\label{Model and Algorithms}%
In this section, we formally define the state, action and reward for the considered scenario. The objective is to determine the HO decisions for any arbitrary waypoints along a given route using a DQN framework.
In Table \ref{DefinitionsModel}, we list the main parameters used in the proposed HO optimization framework.
\begin{table}[!t]
\caption{Definitions in our model related to RL.}
\resizebox{\columnwidth}{!}{
	\begin{tabular}{@{}ll@{}}
		\toprule
		Label & Definition \\ \midrule
		$\mathds{I}(HO)$    &       HO cost   \\
		$w_{HO}$    &       Weight for HO cost   \\
		$w_{RSRP}$    &       Weight for serving cell RSRP   \\
		$s$ &        State defined as $[x_s,y_s,\theta_s,c_s]$     \\   
		$(x_s, y_s)$ &       Position coordinate at state $s$    \\ 
		$\theta_s$ &        Movement direction at state $s$     \\ 
		$c_s$ &        Serving cell at state $s$     \\ 
		$s'$ &        Next state of $s$     \\ 
		$a_s$ &       Action performed at state $s$     \\    
		$a_s'=a_{s'}$ &       Action performed at state $s'$    \\
		$R_{s, a_s}$    &       Reward for taking action $a_s$ in state $s$  \\
		$\mathcal{T}_h$ & Threshold for beginning Q-value iteration\\
		$Q(s,a;w)$   &    Q-value of taking action $a$ at state $s$ (updated at every step)    \\
		$\mathcal{T}_c$& Update cycle for $\hat{Q}(s,a;w)$  \\
		$\hat{Q}(s,a;w)$   &    Q-value of taking action $a$ at state $s$ (updated every $\mathcal{T}_c$ steps)    \\
		$\alpha$    &       Learning rate     \\
		$\lambda$   &       Discount factor  \\
		$\epsilon$   &       Exploration coefficient  \\
		$n$ &Number of training episodes \\
		$ {\boldsymbol{\beta}} $& Replay batch for DQN training\\
		${\boldsymbol{\beta}}_m$ & Minibatch from ${\boldsymbol{\beta}}$ of size $m$\\
		$\mathcal{T}$ &Number of training steps per episode\\
		
		\bottomrule
\end{tabular}}

\label{DefinitionsModel}
\end{table}
\subsection{Definitions}
\textbf{State:} The state of a drone represented by $s= [x_s,y_s,\theta_s,c_s]$ consists of the drone's position $\mathcal{P}_s:(x_s,y_s)$, its movement direction $\theta_s\in \{z\pi/4, z\in \mathbb{Z}, z=0, 1,...,7\}$, and the currently connected cell $c_s\in C$, where $C$ is the set of all candidate cells. We use superscript $'$ to denote the next state $s'$ of a state $s$.
We clarify that the direction of movement $\theta_s$ is restricted to a finite set only for the training phase. For the testing phase, the deep network may output results for other directions, which is beneficial for trajectory adjustment in practical applications.
We describe how a drone trajectory is generated in our model given an initial location $\mathcal{P}_o$ and a final location $\mathcal{P}_e$ of the drone.
At the initial location, we select the movement direction which results in the shortest path to the final location. The drone moves in the selected direction for a fixed distance until it reaches the next waypoint. The same procedure is repeated for selecting direction at each waypoint until it reaches closest to the final location. We note that the resulting drone trajectory is not necessarily a straight line due to a finite number of possible movement directions in our model. We recall that the RL-based HO algorithm merely expects that the drone trajectory is known beforehand. Thus, we are able to get sufficient training data along the route for the DQN. While it is not critical how the fixed trajectories are generated, we have nevertheless described the methodology for the sake of completeness. 

\begin{figure}[!t]
\centering
\subfloat[Illustration of current state and next state.]{%
	\label{Model_part1}
	\includegraphics[width=\columnwidth]{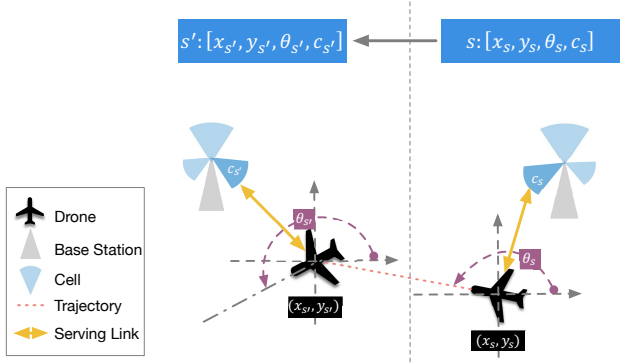}}\hfill
\subfloat[Illustration of action.]{%
	\label{Model_part2}
	\includegraphics[width=.8\columnwidth]{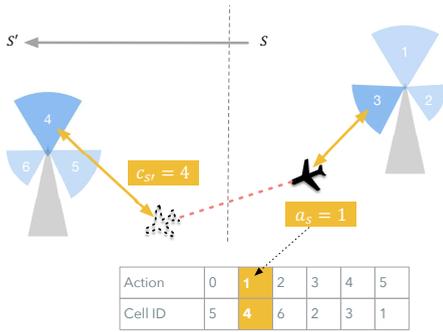}}

\subfloat[Illustration of HO decisions for different waypoints along the flight route.]{%
	\label{Model}
	\includegraphics[width=.8\columnwidth]{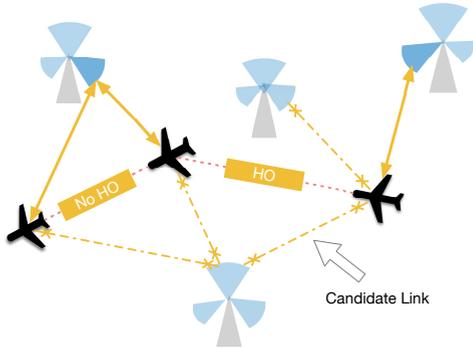}}\hfill 
\caption{Illustration of the proposed RL-based framework.}
\label{Model_parts}
\end{figure}

\textbf{Action:} As the drone trajectory is fixed and known beforehand, the drone position at the future state $s'$ is known a priori at the current state $s$. Therefore, the RSRP values from various cells for the drone position at state $s'$ are also known a priori at state $s$. For the current state $s$, we let $C_{k_s}$ denote the set of candidate cells at the \textit{future} state $s'$, where $C_{k_s}$ consists of the $k$ strongest cells at state $s'$. We assume that the cells in $C_{k_s}$ are sorted in descending order of RSRP magnitudes. The drone's action $a_s$ at the current state $s$ corresponds to choosing a serving cell from $C_{k_s}$ for the next state $s'$. This is illustrated in Fig. \ref{Model_part2} for $k=6$ where the current state is shown by a dashed-line drone and the future state by a solid drone. The 6 cells are sorted in descending order of RSRP magnitudes seen at the future state $s'$, i.e., cell 5 has the largest while cell 1 has the smallest RSRP. The drone takes an action $a_s=1$ at the current state, meaning that it will connect to cell 4 at the next state, i.e., $c_{s'} = 4$. Thus, the action consists of choosing an index (i.e., picking an element) from $C_{k_s}$ at state $s$. As a result, the drone connects to the cell corresponding to that index in state $s'$. 

\textbf{Reward:} We now describe the reward function used in our model.
The goal is to encourage the drone to reduce the number of HOs along the trajectory while also maintaining reliable connectivity. 
In the context of Fig. \ref{Model_part2}, this means that action $0$ (i.e., cell with highest RSRP) is not necessarily always selected. The drone might as well connect to a cell with a lower RSRP at one waypoint that results in fewer HOs at subsequent waypoints. In view of these conflicting goals, we incorporate a weighted combination of the HO cost and the serving cell RSRP at future state in the reward function
\begin{align}
R_{s,a_s}=-w_{HO}\times \mathds{I}(HO)+w_{RSRP}\times RSRP_{s'},
\end{align}
where $w_{HO}$  and $w_{RSRP}$ respectively denote the weights for the HO cost and the serving cell RSRP $RSRP_{s'}$ at state $s'$, while $\mathds{I}(HO)$ is the indicator function for the HO cost such that $\mathds{I}(HO)=1$ when the serving cells at states $s$ and $s'$ are different and $\mathds{I}(HO)=0$ otherwise.
\subsection{Algorithm of HO scheme using DQN}
For complexity reduction, the action space $\mathcal{A}$ in our model is restricted to the strongest $k$ candidate cells for every state. Let us define a set $\mathcal{I}_k=\{0,1,\cdots,k-1\}$ and assume that the trajectory has $l$ waypoints. Unlike using a Q-table \cite{chen2019efficient} to store the Q-values which may require a substantial memory, the Q-value $Q(s,a;w)$ for each state-action pair can be directly obtained from the DQN \cite{mnih2013playing}. 
We describe the training process in Algorithm \ref{algorithm_Q}.
The algorithm complexity is ${\rm O}(\mathcal{T} n)$, where $\mathcal{T}$ is the number of training steps per episode and $n$ is the total number of training episodes. We use two networks for training: one for the initial update of parameter $w$ while the other for storing the more stable $w'$ after $w$ has been appropriately trained for a given period of time.

The Q-value iterations for each training episode are performed in line 5-32. 
An $\epsilon$-greedy exploration is performed in line 9-13 \cite{sutton1998introduction}. The data for each training step is stored in a replay batch $\boldsymbol{\beta}$. Specifically, each row of $\boldsymbol{\beta}$ contains the tuple $(s, a_s, R_{s,s'}, s')$, i.e., current state, action, future state and reward for a training step (line 9-15). 
The training process is activated after $\boldsymbol{\beta}$ has accumulated at least $m$ row entries (line 18). Then, a minibatch $\boldsymbol{\beta}_m$ is obtained by (uniformly) randomly extracting $m$ rows from $\boldsymbol{\beta}$. We let $\boldsymbol{\beta}_{m,S}=[\boldsymbol{\beta}_{m,S}(0), \cdots, \boldsymbol{\beta}_{m,S}(m-1)]$ denote the input state vector consisting of only the current states for all entries in $\boldsymbol{\beta}_m$, where $\boldsymbol{\beta}_{m,S}(i)$ denotes the current state for a row $i\in\mathcal{I}_m$ in $\boldsymbol{\beta}_m$. We feed the input state vector $\boldsymbol{\beta}_{m,S}$ to DQN to compute the Q-values for all possible actions. 
For each state $\boldsymbol{\beta}_{m,S}(i)$, we represent the Q-values for all possible actions by a $k\times 1$ vector $\boldsymbol{Q}(\boldsymbol{\beta}_{m,S}(i);w)=\left[Q(\boldsymbol{\beta}_{m,S}(i),0;w),...,Q(\boldsymbol{\beta}_{m,S}(i),k-1;w)\right]^T$ (line 20). We further define an $m\times k$ matrix
$\mathbb{{Q}}(\boldsymbol{\beta}_{m,S};w) = \left[\boldsymbol{Q}(\boldsymbol{\beta}_{m,S}(0);w),...,\boldsymbol{Q}(\boldsymbol{\beta}_{m,S}(m-1);w)\right]^T$.
As described in line 21-25, we update the entries of $\mathbb{{Q}}(\boldsymbol{\beta}_{m,S};w)$. 
During the preliminary training phase, we update the Q-values using corresponding rewards such that the network has a rough approximation of the Q-values for various state-action pairs. This is because initially the network cannot accurately predict the Q-values used for value iteration. This helps avoid error accumulation in the initial training stage. Then, after running a sufficient number of training steps,
we use the Q-values output from the network for value iteration, as shown in line 21-22. Specifically, we set parameter $\mathcal{T}_h=0.3$ in our model meaning that reward function is used for value iteration for around 30\% steps in each episode, whereas the Q-values are used for the remaining steps.
In this way, the trained parameter $w$ requires only a few oscillations to converge. In addition, the  parameter $w'$ for the target network is updated every $\mathcal{T}_c$ steps (line 25-27), which ensures that $w'$ is replaced by a relatively stable $w$ calculated during the preceding $\mathcal{T}_c$ steps.
Finally, the well-trained target network is used for action prediction for the states along the route. The output from the network is a vector of Q-values for all the possible actions. The action with the highest Q-value is chosen for each state. 

\section{Simulation Results}\label{Experiments and Results}
In this section, we present the simulation results for the proposed DQN-based HO mechanism. For performance comparison, we consider a greedy HO scheme as the baseline in which the drone always connects to the strongest cell. We also contrast the results with those reported in \cite{chen2019efficient} using a tabular Q-learning framework.
We now define a performance metric called the \textit{HO ratio}: for a given flight trajectory, \textit{HO ratio} is the ratio of the number of HOs using the proposed scheme to that using the baseline scheme. By definition, the HO ratio is always $1$ for the baseline case. To illustrate the interplay between the number of HOs and the observed RSRP values, we evaluate the performance for various weight combinations of $w_{HO}$ and $w_{RSRP}$ in the reward function. 
By increasing the ratio ${w_{HO}}/{w_{RSRP}}$, the number of HOs for the DQN-based scheme can be decreased which yields a smaller HO ratio.

\subsection{Data Pre-processing}
\begin{figure}[!t]
\centering
\includegraphics[width=0.95\columnwidth]{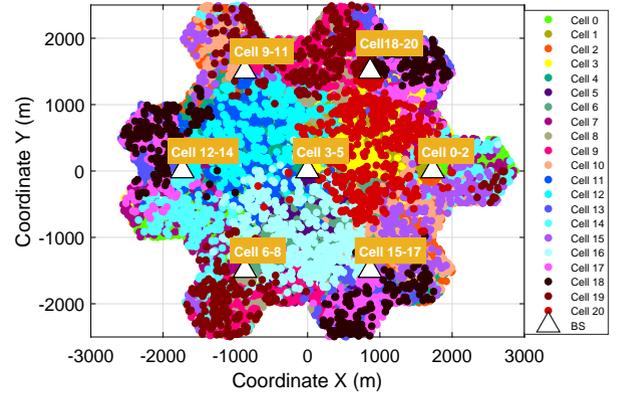}
\caption{Cell association map for the simulated region based on strongest RSRP without quantization \cite{chen2019efficient}.}
\label{samples}
\end{figure}

\begin{figure}
\centering	
\includegraphics[width=1\columnwidth]{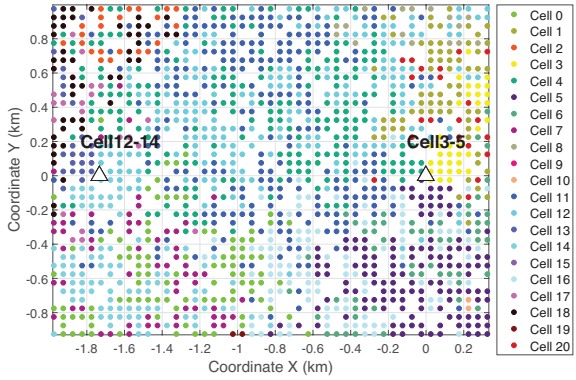}	
\caption{An illustration of the cell association map (for a chunk of the simulated region) based on RSRP  data after quantization \cite{chen2019efficient}.}
\label{quansamples}
\end{figure}

Similar to \cite{chen2019efficient}, we consider a deployment of $7$ BSs in a 2D geographical area of $5\times 6$ km$^2$ where each BS has $3$ cells. We assume that the UEs are located in a 2D plane at an altitude of $50$ m. We generate $10000$ samples of RSRP values for each of these $21$ cells at different UE locations. For normalization, the RSRP samples thus obtained are linearly transformed  to the interval [0 1].
To further quantize the considered 
space, we partition the area into bins of size $50\times 50$ m$^2$ (as shown in Fig. \ref{samples} and Fig. \ref{quansamples}). For each bin, we compute the representative RSRP value for a cell as the average of the RSRP samples in that bin. 

\subsection{Experimental Setup}\label{Results using Q-learning}
\begin{figure}[!t]
\centering
\includegraphics[width=1\columnwidth]{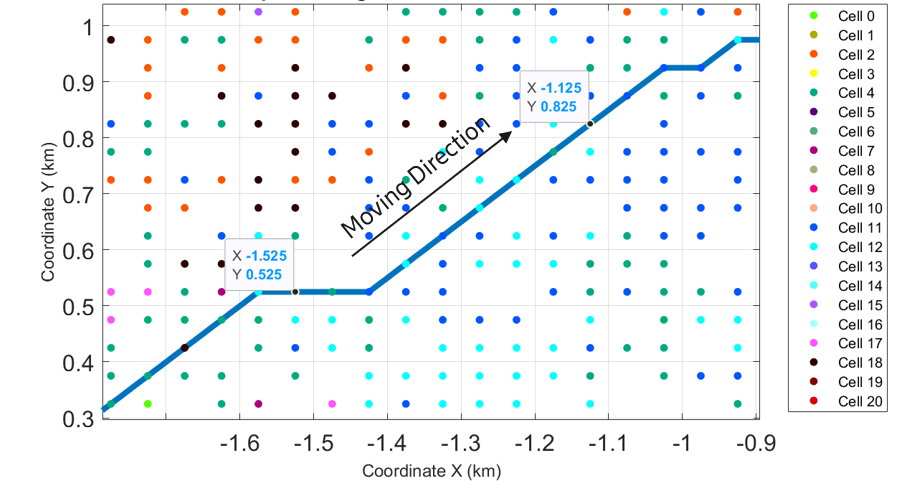}
\caption{A zoomed portion of a drone's route along with the serving cells at each waypoint \cite{chen2019efficient}.}
\label{temp_traj}
\end{figure}

We simulate the performance using $2000$ runs for each of the DQN-based, the Q-learning-based \cite{chen2019efficient} and the baseline schemes. 
For each run, the testing route is generated randomly as explained in Section \ref{Model and Algorithms}. We show a snapshot of a flying trajectory in Fig. \ref{temp_traj}. The distance between subsequent waypoints along the trajectory is set to 50 m. We note that the drone's speed is not relevant since we aim to reduce the number of HOs for a given trajectory rather than the number of  HOs per unit time.   
For the DQN-based scheme, we use a neural network with two fully-connected hidden layers and train it using RMSprop as the optimizer. We use the following parameter values for Q-value iteration: $n=120$, $\mathcal{T}=1000$,  $\mathcal{T}_c=20$, $\mathcal{T}_h = 0.3$, $\lambda=0.3$, $\epsilon=0.2$ and $m=64$.

\begin{figure}[!t]
\centering
{%
	\includegraphics[width=\columnwidth]{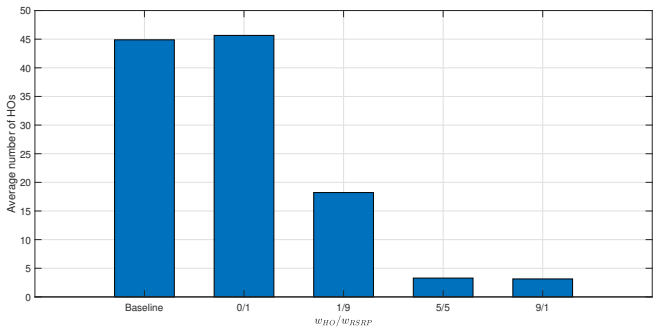}
}
\caption{Average number of HOs for various weight combinations ($w_{HO}/w_{RSRP}$).}\label{ab_Q}
\end{figure}

\begin{figure}[!t]
\centering
{%
		\includegraphics[width=\columnwidth]{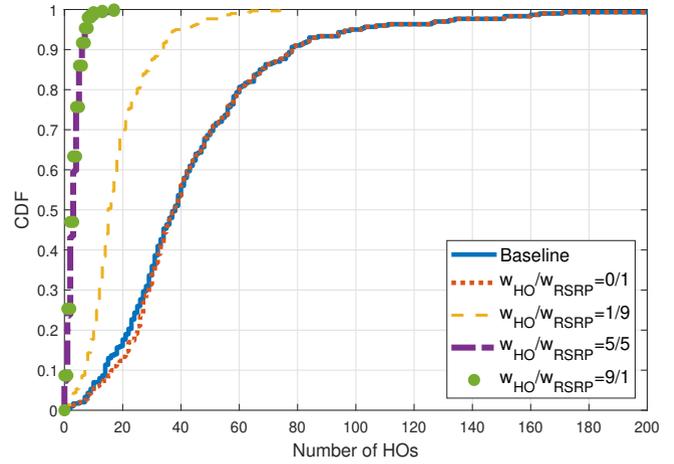}
}
\caption{CDF of the number of HOs.}\label{CDF_HO_Q}
\end{figure}

\begin{figure}[!t]
\centering
{%
	\includegraphics[width=\columnwidth]{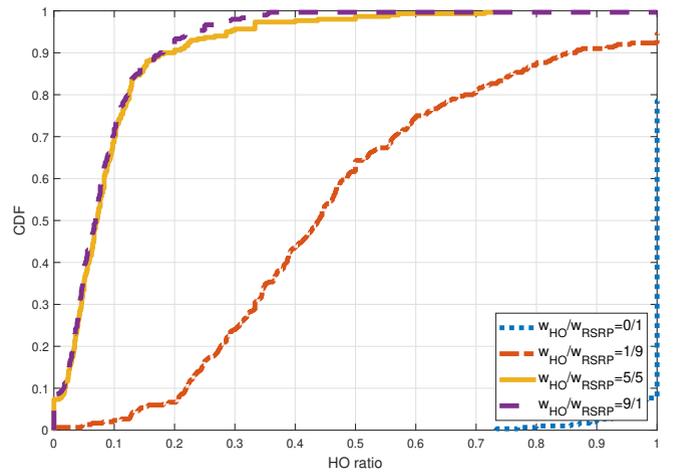}
}
\caption{CDF of the HO ratio.} \label{CDF_HOratio_Q} 
\end{figure}

\begin{figure}[!t]
\centering
\includegraphics[width=\columnwidth]{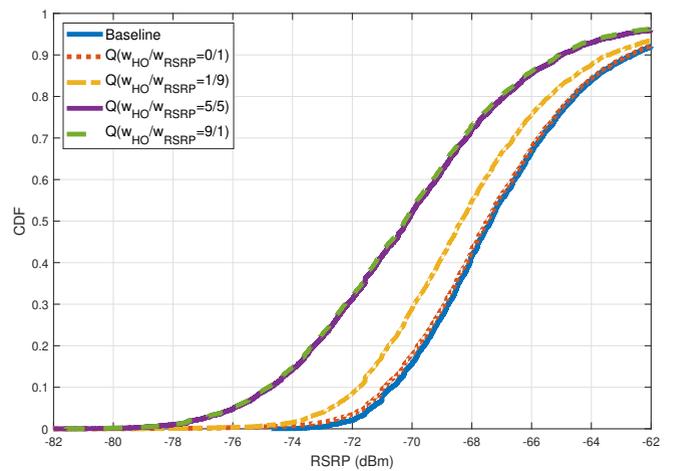} 
\caption{CDF of RSRP for various weight combinations.}
\label{CDF_RSRP_Q}
\end{figure}

\subsection{Results}
In Fig. \ref{ab_Q}, we plot the average number of HOs per flight for various weight combinations. We first consider the case of practical interest where the HO cost is non-zero. The proposed approach helps avoid unnecessary HOs compared to the baseline case even for a modest weight for the HO cost. 
For example, introducing a slight penalty for a HO event by setting ${w_{HO}}/{w_{RSRP}}={1}/{9}$ helps cut the average number of per-flight HOs roughly in half.
By further increasing ${w_{HO}}/{w_{RSRP}}$, the HO cost increases which reduces the number of HOs. For instance, the number of HOs are reduced by around 11 times when ${w_{HO}}/{w_{RSRP}}=1$. We further note that there are diminishing returns if the HO cost is weighed higher than the RSRP, i.e., when ${w_{HO}}/{w_{RSRP}}>1$. 
We now consider the case where there is no HO cost, i.e., $w_{HO}=0$. The proposed scheme performs slightly worse than the baseline in terms of the average number of per-flight HOs. This apparent anomaly is because the Q-value obtained from a DQN-based algorithm is in fact an approximation of that obtained via tabular Q-learning. As reported in \cite{chen2019efficient}, the case $w_{HO}=0$ is equivalent to the baseline when tabular Q-learning is used. Nonetheless, we note that this corner case ($w_{HO}=0$) is irrelevant as the network can revert to the baseline HO approach instead.

In Fig. \ref{CDF_HO_Q}, we plot the cumulative distribution function (CDF) of the number of per-flight HOs. For a non-zero HO cost, the proposed scheme significantly reduces the number of HOs. For example, with a probability of 0.95, the number of per-flight HOs is expected to be fewer than 98 
for the baseline case. For the same probability, the proposed scheme reduces the number of HOs to fewer than 38
 for ${w_{HO}}/{w_{RSRP}}={1}/{9}$ and to fewer than 7 for ${w_{HO}}/{w_{RSRP}}=1$. Similarly, with a probability of 0.1, fewer than 14
 per-flight HOs are expected for the baseline case. The proposed scheme requires fewer than 7 HOs for ${w_{HO}}/{w_{RSRP}}={1}/{9}$ and only 1 HO for ${w_{HO}}/{w_{RSRP}}=1$.
For the special case $w_{HO}=0$, we observe that the CDF for the proposed approach is slightly worse than that of the baseline. This trend is consistent with the explanation provided for Fig. \ref{ab_Q}. 

We caution that merely inspecting the absolute number of HOs may be misleading as it does not reflect the reduction in HOs on a per-flight basis.   
In Fig. \ref{CDF_HOratio_Q}, we plot the CDF of the HO ratio for the proposed scheme. We note that this metric captures the reduction in number of HOs relative to the baseline for \textit{each} flight. For ${w_{HO}}/{w_{RSRP}}={1}$, a HO ratio of 0.1, 0.2 or 0.5 can be achieved with a probability of 
0.70, 0.90 or 0.98, respectively. This means a reduction in the number of HOs by \textit{at least} 2 times for 98\% flights, 5 times for 90\% flights, or 10 times for 70\% flights.  
In short, by properly adjusting the weights for the HO cost and RSRP, the RL-based scheme can significantly reduce the number of HOs for various scenarios.

In Fig. \ref{CDF_RSRP_Q}, we plot the CDF of the RSRP observed along the trajectory of the drone UE for various combinations of HO cost and RSRP weights. 
We note that the proposed scheme provides a flexible way to reduce ping-pong HOs (and resulting signalling overheads) while sacrificing RSRP. 
For example, when ${w_{HO}}/{w_{RSRP}}=1$, a (worst-case) 5th-percentile UE suffers an RSRP loss of around 4.5 dB relative to the baseline. If such degradation is not acceptable, setting ${w_{HO}}/{w_{RSRP}}={1}/{9}$ will incur only a meager loss in RSRP.
It is evident from Fig. \ref{CDF_HO_Q} and Fig. \ref{CDF_RSRP_Q} that both choices substantially reduce the number of HOs compared to the baseline. 
We remark that the operating conditions will influence the network's decision to strike a favorable tradeoff between the HO overheads and reliable connectivity.
In Fig. \ref{CDF_RSRP_Q}, the minimum RSRP exceeds -82 dBm which translates to a signal-to-noise ratio (SNR) of 31 dB assuming a bandwidth of 1 MHz and a noise power of -113 dBm, which is usually sufficient to provide reliable connectivity.

\subsection{Comparison with Q-learning based approach}
Let us compare the performance in terms of the HO ratio and RSRP for the schemes based on DQN and Q-learning \cite{chen2019efficient}.
As evident from Table \ref{table: compare}, the Q-learning-based approach \cite{chen2019efficient} yields a smaller HO ratio than that based on DQN. As noted previously, this is because the DQN attempts to approximate the Q-values obtained via tabular Q-learning \cite{chen2019efficient}. 
In Table \ref{table: compare rsrp}, we
include some selected points from the RSRP CDFs for both cases. We observe only a negligible drop in RSRP for the DQN-based scheme compared to the Q-learning approach.  
Despite the performance differences, both RL-based methods can significantly reduce the number of HOs while maintaining reliable connectivity. Furthermore, as a first step, we considered 2D drone mobility in a rather limited geographical area. In practical scenarios with longer flying routes, the state space may grow prohibitively large with an approach based on tabular Q-learning. This renders the proposed DQN-based method more appealing thanks to reduced implementation complexity.

\begin{table}[!t]
\centering
\caption{A comparison of HO ratio CDFs of Q-learning \cite{chen2019efficient} and DQN for various weight combinations.}
\resizebox{7.5cm}{!}{ 
		\begin{tabular}{|c|c|l|l|l|}
			
			\hline
			{Method}  & {probability}  &\multicolumn{2}{r}{HO Ratio} & 
			\\ 		\hline
			\multicolumn{2}{|c|}{$w_{HO}/w_{RSRP}$} & 0/1 & 1/9 & 5/5  
			\\\hline
			\multirow{4}{*}{Q-learning}             & 0.05  & 1        & 0.130 & 0  \\ 
			& 0.50 & 1        & 0.314 & 0.064 
			\\
			& 0.95 & 1        & 0.579 & 0.171 \\
			\hline
			\multirow{4}{*}{DQN}           & 0.05  & 0.96    & 0.143 & 0
			\\
			& 0.50 & 1        & 0.439 & 0.071 \\
			& 0.95 & 1.389    & 1.056 & 0.286 \\
			\hline
		\end{tabular}
}
\label{table: compare}
\end{table}

\begin{table}[!t]
	\centering
	\caption{A comparison of RSRP CDFs of Q-learning \cite{chen2019efficient} and DQN for various weight combinations.}
	\resizebox{7.5cm}{!}{ 	
			\begin{tabular}{|c|c|l|l|l|}
				
				\hline
				{Method}  & {probability}  &\multicolumn{2}{r}{RSRP (dBm)} & 
				\\ 		\hline
				\multicolumn{2}{|c|}{$w_{HO}/w_{RSRP}$} & 0/1 & 1/9 & 5/5  
				\\\hline
				\multirow{4}{*}{Q-learning}             & 0.05  & -71.33        & -72.88 & -76  \\ 
				& 0.50 & -67.25        & -67.92 & -70.1 
				\\
				& 0.90 & -62.69      & -62.93 & -64.96 \\
				\hline
				\multirow{4}{*}{DQN}           & 0.05  & -71.6    & -72.73 & -76
				\\
				& 0.50 & -67.35     & -68.35 & -70.1 \\
				& 0.90 & -62.70    & -63.43 & -65 \\
				\hline
			\end{tabular}
	}
	\label{table: compare rsrp}
\end{table}

\section{Conclusion}
\label{Conclusion}
In this paper, we have developed a novel deep RL-based HO scheme to provide efficient mobility support for a drone served by a cellular network. 
By exploiting a DQN approach, we have proposed a flexible mechanism for dynamic HO decision making based on the  drone’s flight path and the distribution of the BSs.
We have shown that the proposed HO mechanism enables the network to manage the tradeoff between the number of HOs (i.e., overheads) and the received signal strength by appropriately adjusting the reward function in the deep RL framework. 
 The results have demonstrated that, compared to the greedy HO approach where the drone always connects to the strongest cell, the deep RL-based HO scheme can significantly reduce the number of HOs while maintaining reliable connectivity. %

There are several potential directions for future work. 
A natural extension will be to include 3D drone mobility in the current framework.
It will also be worth 
validating the proposed scheme for larger testing areas and/or longer flying trajectories with a larger pool of candidate cells. 
Another notable contribution will be to enhance the model with additional parameters to account for inter-cell interference. %




\footnotesize
\bibliographystyle{IEEEtran}
\bibliography{IEEEabrv,Main_v3}

\EOD

\end{document}